\documentclass[12pt]{article}
\usepackage{latexsym,epsfig,fancyhdr}

\newcommand{\project}{GRaNDScan~}               
\newcommand{\projectc}{GRaNDScan}               

\begin{document}

\title{Conceptual Design of a Cosmic Ray Detector
Operating Between $10^{17}$ and $10^{19}$\,eV for the Study of the
Galactic Center} 
\author{T.\,Adams$^{1}$,  E.C.\,Loh$^{2}$, 
S.\,BenZvi$^{3}$, S.\,Westerhoff$^{3}$\\
\it\footnotesize $^{1}$ Florida State University, Department of Physics,
Tallahassee, Florida, USA\\
\it\footnotesize $^{2}$ University of Utah, Department of Physics,
Salt Lake City, Utah, USA\\ 
\it\footnotesize $^{3}$ Columbia University, Department of Physics, New York, New York, USA}

\maketitle

\begin{abstract}
For our understanding of the origin of ultra high energy cosmic
rays, the energy region between $10^{17}$ and $10^{19}$\,eV is 
of crucial importance.  Previous experiments have found indirect 
evidence that at these energies, the origin of cosmic rays changes 
from predominantly Galactic to extragalactic.  In addition, weak
evidence for an excess of cosmic rays from the direction of the
Galactic center in a narrow energy band around $10^{18}$\,eV 
has been claimed.

However, so far there is no direct evidence supporting this scenario.
Neither Galactic nor extragalactic sources have been unambiguously
established.  Given the importance of this energy range, there is
a strong case for a dedicated experiment to study the EeV energy
region with high precision.

We present the conceptual design of \projectc, a mobile stereo 
air fluorescence detector optimized to study the energy spectrum, 
composition, and arrival direction of cosmic rays in this important 
energy range.  If located at a site on the southern hemisphere 
with good exposure to the Galactic center, this type of experiment 
will provide an accurate map of the Galactic center region, long 
suspected to harbor one or several sources of ultra high energy 
cosmic rays.
\end{abstract}

\noindent------------------------------------------------------\\
\footnotesize
$^1$\texttt{tadams@hep.fsu.edu}\\
$^2$\texttt{ecloh@umdgrb.umd.edu}\\
$^3$\texttt{sybenzvi@nevis.columbia.edu}\\
$^4$\texttt{westerhoff@nevis.columbia.edu}
\normalsize

\newpage
\section{Introduction}\label{s:intro}

The goal of \projectc\footnote{{\it G}amma {\it R}ay {\it a}nd 
{\it N}eutron {\it D}ecay {\it Scan} of the Galaxy} is to study 
the energy spectrum, chemical composition, and arrival direction 
of cosmic rays in the energy range from $10^{17}$ to $10^{19}$\,eV 
with high sensitivity from a site with visibility of the Galactic 
center region.

The case for this study is strong in several respects.  By
concentrating on this energy range, \project will provide
data of unprecedented quality in a region where
\begin{itemize}
\item{the cosmic ray energy spectrum shows
features, the `ankle' and (less prominent) the `second knee,'} 
\item{the chemical composition undergoes an important change 
from a heavier to a lighter mixture,}  
\item{a cosmic ray flux enhancement from the region around 
the Galactic center has been claimed.}  
\end{itemize}
Most of these features have not been studied with a dedicated
instrument, and consequently their statistical significance 
is unsatisfying at this point.   With the wealth of information
that this energy range offers, this study will 
provide crucial information on the origin of ultra high energy 
cosmic rays and the acceleration mechanisms at work.

In our current understanding, the changes in composition 
and energy spectrum are indicative of a transition in the 
nature of the cosmic ray origin itself.  Whereas cosmic 
rays below $10^{18}$\,eV are mostly Galactic in origin, a new
extragalactic component becomes dominant at higher energies.

Unfortunately, there is no {\it direct} evidence that
supports this general picture -- neither Galactic sources
at energies below the ankle nor extragalactic sources at
higher energies have been unambiguously detected.  The most 
likely acceleration site in our own Galaxy is the region 
around the Galactic center, which stands out as its most 
energetic region.  Radio, far infrared and $\gamma$-ray data
indicate that the star formation and supernova activity of 
our Galaxy peaks in the center.  This general picture is 
confirmed by studies of other disk galaxies.  Correlations 
of data at radio and infrared wavelengths suggest that the 
cosmic ray production is generally higher in galactic center 
regions, just as the star formation rate is higher.

We therefore propose to operate \project from a site on the 
southern hemisphere, with good visibility of the center region.
This location will enable us to combine the general 
study of the composition and the energy spectrum with a 
detailed analysis and mapping of the Galactic center region,

Earlier results from AGASA~\cite{bib:agasa}, SUGAR~\cite{bib:sugar}, 
and Fly's Eye~\cite{bib:flyseye} indicate that the Galactic 
center region may indeed harbor one or several sources of 
cosmic rays at $10^{18}$\,eV.
However, the statistical significance of the results is poor, 
mainly because these experiments were not optimized for the study 
of the Galactic center region in this energy range.  An important
goal of \project is to clarify this unsatisfying situation and 
establish or disprove claims of an enhanced flux from the Galactic 
center with high significance.

The AGASA and SUGAR results naturally raise the question of the 
chemical nature of the cosmic ray flux from sources inside our 
Galaxy.  While protons cannot reach us without deflection, it is well known 
that neutrons with energy 1\,EeV can traverse the Galactic field 
undisturbed and reach us un-decayed from the Galactic center, a 
mere 8 kpc from the solar system.  With higher energies, neutrons 
could reach us from anywhere in our Galaxy.  In short, a properly 
designed detector could use neutrons as a tool for performing 
tomographic searches for sources of cosmic rays in our Galaxy.

The basic requirements for a multi-purpose detector like
\project are therefore excellent energy resolution, good angular
resolution, and the ability to discriminate between different
primaries, mainly $\gamma$'s, hadrons, and heavier nuclei
like iron.  All composition studies rely heavily on the quality
of our understanding of the first interaction of cosmic ray primaries in the
atmosphere and the development of the shower cascade.  However,
this dependence can be minimized with a detector that
{\it directly} observes the development of the shower cascade.  
The air fluorescence technique meets all these requirements,
and \project is designed as an air fluorescence detector
in the tradition of the Fly's Eye and the HiRes experiment. 

Taking the SUGAR detector as a baseline detector, \project is 
designed with a factor of ten better angular resolution, a factor 
of ten improved energy resolution, more than a factor of ten larger 
aperture with shower profile measuring capability, a capability 
that SUGAR did not have.  The profile measuring capability would 
enable the \project detector to identify neutrons from a source 
by examining the mass composition of cosmic ray from the source 
and the mass composition of off-source cosmic rays.  Similarly, 
$\gamma$-rays can also be identified by similar profile comparisons.  
The improved energy resolution would also allow one to compare 
on-source and off-source cosmic spectrum and thus to understand 
how cosmic rays are accelerated.

It should be noted that except for the Pierre Auger 
Array~\cite{bib:auger1}, there are no high energy cosmic ray 
detectors in the southern hemisphere.  The main focus of the Auger 
Array is the study of cosmic rays above $10^{19}$\,eV, 
and the layout of the Auger air fluorescence detectors is not optimized
for studying cosmic rays far below this threshold.  Using events 
that trigger a single fluorescence eye and one or more ground 
detectors, Auger can extend the sensitive range down to 
$10^{18}$\,eV~\cite{bib:auger2}, but not far below that, and without 
the advantages of the stereo fluorescence technique.

In the following sections, we will first review some of the 
experimental findings (Section\,\ref{s:experiments}), then
analyze the theoretical problems raised by the data 
(Section\,\ref{s:theory}).  Section\,\ref{s:goals} outlines the
scientific goals of the \project project, and 
Section\,\ref{s:design} discusses the detector design. 

\section{Experimental Results}\label{s:experiments}

\subsection{Energy Spectrum and Composition}

\begin{figure}
\begin{center}
\epsfig{file=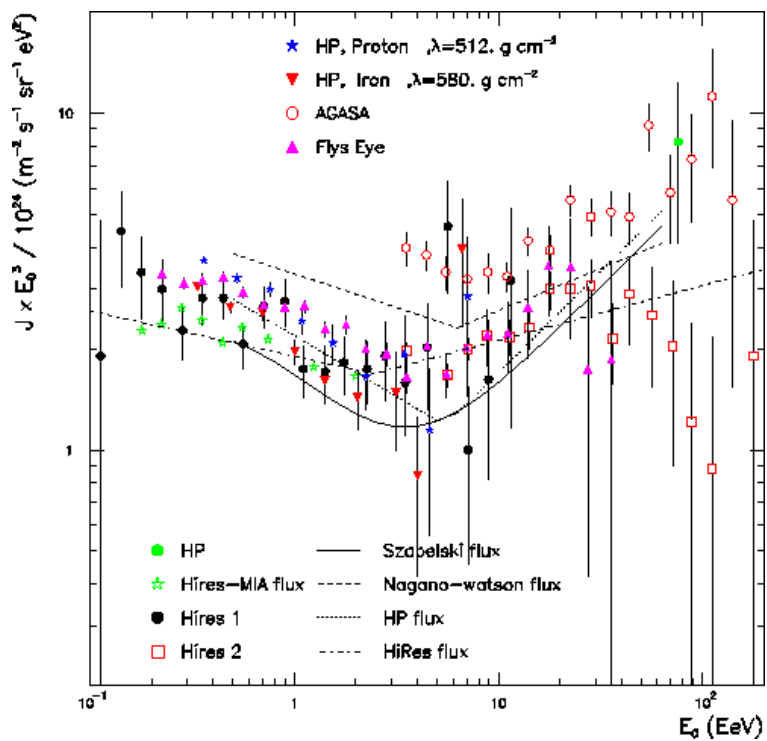,width=10cm}
\caption{\protect\sl Summary of measurements of the cosmic ray
energy spectrum above $10^{17}$\,eV for various experiments.
The flux is multiplied by $E^{3}$ to enhance spectral 
features.  Taken from~\cite{bib:anchordoqui}.}
\label{fig:spectrum}
\end{center}
\end{figure}

There are two prominent breaks in the cosmic ray energy
spectrum, a steepening around $5\cdot 10^{15}$\,eV
(the `knee') and a flattening around $3\cdot10^{18}$\,eV
(the `ankle').  The latter can be seen in Fig.\,\ref{fig:spectrum},
a summary plot of the current experimental status of cosmic ray energy
spectrum measurements above $10^{17}$\,eV~\cite{bib:anchordoqui}.
The knee and the ankle are correlated with changes
in the average chemical composition of the cosmic ray flux.
The KASCADE~\cite{bib:kascade} experiment has shown that 
the composition is mainly heavy (iron) at energies 
above the cosmic ray ``knee'' at $10^{15}$\,eV.
Around the ankle, data from the Fly's Eye and HiRes-MIA
hybrid experiment indicate a change to a proton-dominated
composition.  

Both Fly's Eye and its successor HiRes (High Resolution 
Fly's Eye) are air fluorescence detectors,
operated in the clean atmosphere of the Utah desert. 
Air fluorescence detectors observe the longitudinal
shower profile and deduce the chemical composition
using the fact that showers induced by lighter particles 
penetrate more deeply into the atmosphere.  Due to
large intrinsic fluctuations in the shower development,
the determination of the particle type is not possible
on an event-by-event basis, but the rate of change
of the {\it average} shower maximum per logarithmic decade,
the `elongation rate,' is a good indicator of the
chemical composition.  Results from Fly's Eye~\cite{bib:fe_comp} 
and the HiRes-MIA hybrid detector~\cite{bib:chihwa,bib:hires}
suggest that near the ankle, the average mass composition 
changes from iron-dominated to proton-dominated, and 
stays consistently proton-dominated at higher energies. 

Both results lend credibility to a scenario that attributes 
the dominant cosmic ray flux below and above the ankle to 
two different source populations.  Galactic sources, 
dominating the flux at lower energies, run out of steam 
around the ankle, and extragalactic contributions become 
dominant.

The exact location of the end of the Galactic component
and therefore the maximum energy Galactic sources can
achieve might be slightly below the ankle.
There is marginal evidence in the Haverah Park~\cite{bib:hp},
Fly's Eye~\cite{bib:bird}, AKENO and HiRes/MIA~\cite{bib:hires}
spectrum for a second steepening of the spectrum near $3\cdot 10^{17}$ eV.
Various models claim that this `second knee' caused by the
disappearance of the Galactic component - the heavy elements 
that dominate the all-particle spectrum above the (first) 
knee finally reach the highest possible energy and 
subsequently drop out, causing the spectrum to 
steepen~\cite{bib:biermann1,bib:biermann1a,bib:biermann1b}.  
This means that the location of the second knee provides important 
information on Galactic acceleration mechanisms.

\subsection{Excess From the Galactic Center}

There is marginal evidence for a cosmic ray source in the vicinity
of the Galactic center.  To date, data from three different 
experiments have been used to search for this anisotropy: AGASA, 
Fly's Eye, and SUGAR.  A summary of the global coordinates and 
detector characteristics is given in Table~\ref{tab:expcomp}.  

In J2000 equatorial coordinates, the Galactic center is located at 
right ascension $\alpha=266.4^{\circ}$ and declination 
$\delta=-28.9^{\circ}$.  AGASA and Fly's Eye, both located on the 
northern hemisphere, observe the Galactic Center near the edge of 
their acceptance. SUGAR is so far the only instrument to observe it
from a location in the southern hemisphere.  All three experiments 
were sensitive to very high energy cosmic rays ($E>10^{17}$\,eV), 
but used different detection methods.  

The AGASA experiment used a $100\,\mathrm{km}^2$ ground based shower 
array located at the Akeno Observatory in Japan operated since 1984.
A harmonic analysis was performed on a sample of 114,000 events passing 
their selection criteria~\cite{bib:agasa}. The first harmonic revealed 
an excess in the region near $10^{18}$\,eV with an amplitude of $4\,\%$.  
Analysis of the energy dependence shows the most significant excess 
falls in the range $10^{17.9}-10^{18.3}$\,eV.
 
\begin{figure}
\begin{center}
\epsfig{file=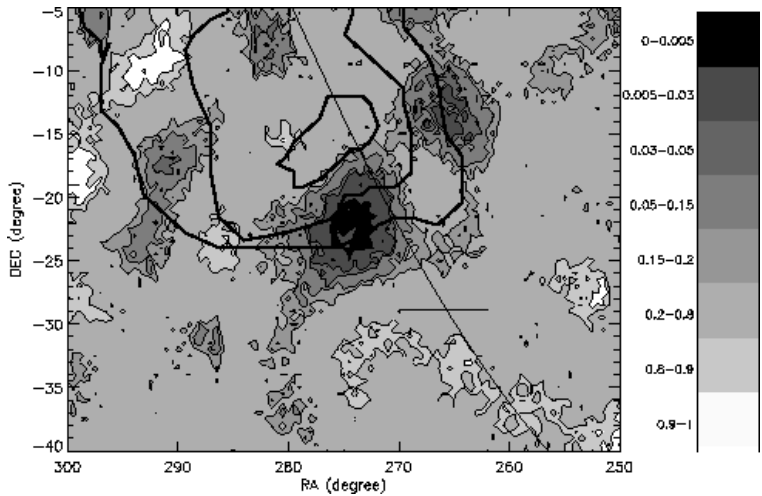,width=10cm}
\caption{\protect\sl Comparison of AGASA and SUGAR results (taken from
\cite{bib:sugar}).  The underlying b/w plot represents the SUGAR results 
in chance probability as a function of right ascension $\alpha$ and 
declination $\delta$,  and the solid lines indicate the $2\,\sigma$, 
$3\,\sigma$, and $4\,\sigma$ contours of the AGASA excess.  The position
of the Galactic center at $(\alpha, \delta)= (266.4^{\circ},-28.9^{\circ})$
is indicated.}
\label{fig:sugar_agasa}
\end{center}
\end{figure}

A sky map of the AGASA data shows a $4\,\sigma$ excess in the direction 
of the Galactic center, a $3\,\sigma$ excess in the Cygnus region and 
a $-3.7\,\sigma$ deficit in the direction of the anti-Galactic center.  
The Galactic center is outside of the visibility of the AGASA detector 
and the excess lies in the Galactic plane at the edge of acceptance.  
Fig.\,\ref{fig:sugar_agasa} shows the AGASA map near the Galactic center.

\begin{table}[t]
\begin{center}
\begin{tabular}{|c||cc|c|c|}
\hline
Experiment & Latitude & Longitude & Detector Type & 
Running Time \\ \hline
AGASA      & $35.8^{\circ}$ N & $138.5^{\circ}$ E &
        ground based array & 1990 - present \\
Fly's Eye  & $40.2^{\circ}$ N & $112.8^{\circ}$ W & 
        air fluorescence  & 1981 - 1993  \\
SUGAR      & $30.5^{\circ}$ S & $148.6^{\circ}$ E & ground muon array  & 
        1968 - 1979 \\ \hline
\end{tabular}
\caption{\label{tab:expcomp} Comparison of characteristics of AGASA,
           Fly's Eye, and SUGAR.}
\end{center}
\end{table}

The Fly's Eye anisotropy analysis~\cite{bib:flyseye} showed an 
enhancement in the Galactic plane which had a probability of 0.006 
for resulting from a fluctuation.  The enhancement was most 
significant in the energy range $0.4-1.0\times 10^{18}$\,eV.  
A latitude gradient was searched for, but was not found to be 
statistically significant ($<2\sigma$).     

The Sydney University Giant Air Shower Recorder (SUGAR) studied very high 
energy cosmic rays during the period 1968-1979.  Located in New South 
Wales, Australia, the experiment measured shower muons $>0.75$\,GeV
using a $70\,\mathrm{km}^2$ ground array~\cite{bib:sugarexp} with 47
liquid scintillation counters buried 1.5\,m deep under ground on a grid
with 1.6\,km spacing.  A denser inner sub-array of $1\,\mathrm{km}^2$ size
with 500\,m grid spacing was sensitive to lower energy showers.

A re-analysis of the SUGAR data by Bellido et al.~\cite{bib:sugar}
determined {\em a priori} the energy range and binning based on the AGASA 
analysis to avoid the trial penalty which makes the 
interpretation of the AGASA results so difficult.  
The analysis found an excess of background probability 0.005
at $(\alpha,\delta) = (274^{\circ},-22^{\circ})$, which is close to but 
significantly $(7.5^{\circ})$ different from the location of the Galactic 
center and does not coincide with the position of the AGASA excess.  
Fig.\,\ref{fig:sugar_agasa} compares the excess in the Galactic center 
region as seen by AGASA and SUGAR.  The positions differ by 
$6^{\circ}$, and while the AGASA excess shows significant smearing, the 
SUGAR excess appears as a point source within the array's resolution.

The statistically rigorous way the SUGAR analysis was performed allows
a first estimate of the flux from the unknown source, which was found to
be $(2.7\pm0.9)\,\mathrm{km}^{-2}\mathrm{yr}^{-1}$ in the energy interval
between $17.9<\mathrm{log}(E/\mathrm{eV})<18.5$.  It should be noted that 
due to the small aperture of the instrument at these low energies, the
accumulated excess amounts to only 10 events in 11 years of running time.

In summary, results from the northern hemisphere 
detectors, Fly's Eye and AGASA, seem to indicate a broad cosmic ray 
enhancement near the Galactic center, while SUGAR 
reports an enhancement consistent with a point source.
With the known Galactic magnetic fields, it is reasonable 
to conclude that the results from the three experiments are 
consistent with each other.  If a source existed near the Galactic 
center, the two detectors in the northern hemisphere, with very 
limited exposure to the Galactic center, would detect mostly 
particles bent by the magnetic fields.  The SUGAR detector, on
the other hand, with a full view of the Galactic center, would 
be able to detect cosmic rays from the source.
Albeit statistically not strong, the experimental results are consistent 
with each other, and they constitute strong and compelling reasons 
for pursuing the source study proposed in this paper.

\section{Theoretical Issues}\label{s:theory}

\subsection{Cosmic Particle Sources and Acceleration}
\label{ss:acceleration}

In our current understanding, cosmic ray particles are 
accelerated to ultra high energies by shock acceleration, 
a process initially proposed by Fermi~\cite{bib:fermi}.  
If shock acceleration is indeed the dominant process, any 
potential source has to meet requirements summarized by 
Hillas~\cite{bib:hillas}.  The source must be able to 
(magnetically) confine the particles during the acceleration 
process, and the source environment must allow particles to 
eventually escape without substantial energy losses such as
photopion production in region of high 
photon density.  This limits the magnetic fields 
in the source vicinity, the source size, and the energy 
density of photons within the source.  At $10^{18}$\,eV, 
source parameters are at the edge of the allowed regions.
Given all these requirements, conditions in known 
Galactic sources can not readily generate cosmic ray energies 
this high.  
Shock acceleration is at work in supernovae exploding into 
the interstellar medium, and in our current understanding,
these supernovae are responsible for the cosmic ray flux
below the knee at about $10^{15}$\,eV.

If the energy spectrum continues above $10^{18}$\,eV in 
violation of the Hillas limit, this might be an indication 
that our understanding of the acceleration process is wrong 
or incomplete, and other more efficient, but yet unknown 
processes, are at work.

A possible source for cosmic rays above $10^{14}$\,eV has been
proposed by Biermann~\cite{bib:biermann1,bib:biermann2,bib:biermann3}.  
The model suggests that these cosmic rays also originate from shock 
acceleration in supernovae, but their sources are supernovae
which explode into their own strong stellar wind rather than 
into the interstellar medium.  Wind supernovae are produced by 
massive stars with initial zero age main sequence masses above 
20 solar masses.  These stars explode near their birthplace, 
where the material from which they were formed is still 
around~\cite{bib:biermann4}.  Wind supernovae have an energy of 
explosion near $10^{52}$\,erg, and since the stellar wind has 
a density gradient, the shock speed stays high even at larger distances
and energies up to $3\,\cdot 10^{18}$\,eV for iron nuclei are 
possible.

This illustrates the impact of the study of cosmic
particle origins at \project energies.  Understanding their origin
may have consequences important for stellar evolution, supernova 
physics, and particle physics.

\subsection{Galactic Magnetic Fields and Charged Particles}
\label{ss:magnetic}

If the primary particles are protons or heavier nuclei, 
we also need to understand Galactic magnetic fields.
Charged cosmic ray particles are subject to large deflections
in ambient magnetic fields unless their Larmor radius is 
substantially larger than the distance between the source 
and the observer.  

At present, the origin of Galactic magnetic fields is not 
well-understood.  The strength of the field in the solar vicinity
is about $6-7\,\mathrm{\mu G}$, with a regular and an irregular component 
both contributing about half the total (see the review and and 
summary in~\cite{bib:beck}).  The regular component points along
the spiral arms inward and is observed in all spiral galaxies 
which have been well-studied~\cite{bib:krause}.  It seems consistent 
with what we know about our own Galaxy.  The irregular component 
is turbulent and possibly a result of the superposition of supernova 
shells.  It is non-uniform with a typical length scale of 100\,pc.  
A $10^{18}$\,eV proton in a field of $\mathrm{\mu G}$ strength has 
a Larmor radius of 300\,pc, while the distance between the Galactic 
center and the solar system is about 8\,kpc.  Charged particles can 
therefore not get from the Galactic center to us on a straight line.  

If we indeed see a charged component, then the AGASA excess might
be explained by diffusion of cosmic rays from the Galactic 
center region past us.
Clay et al.~\cite{bib:clay1} have modeled the propagation of 
charged cosmic ray particles and shown that any gradient in 
the flux of cosmic rays leads to a weak anisotropy and creates
an excess of events which is distributed as a halo around the
source direction.  The estimated halo diameter of $20^{\circ}$ 
is consistent with the AGASA results (although the results are
weak and do not constitute much of a constraint). 
The anisotropy is expected to show a symmetry in opposite 
directions of the sky, and AGASA data in fact indicates a weak 
deficit in the direction of the Galactic anti-center.

As shown in~\cite{bib:clay2}, this diffusion model can also account 
for the relatively small energy region of the excess.
At energies below $10^{17}$\,eV, the propagation is diffuse
and the expected anisotropy is small.  With increasing energy
(above $10^{17}$\,eV for protons and $10^{18.7}$\,eV for iron)
the particles start to retain their directional information to
some degree, and a directional excess with a broad halo around
the source direction becomes visible.  At even higher energies,
the source itself may no longer be able to produce these 
energies, or, if it is, the flux from the source is overshadowed
by other components, for example extragalactic sources.
This would make a detection of the flux 
from the Galactic center statistically challenging with the 
limited amount of data currently available, and the conclusion is
that any excess from the Galactic center is expected
in a narrow energy range only.  

The diffusion should result in a cosinusoidal dependence of 
the flux on angle from the Galactic center. Any future 
detector needs to be sensitive enough to measure the flux 
as a function of angular distance from the center to test 
this important prediction.  The AGASA data, although skewed 
by the fact that the center itself is beyond the field of 
view and the center region is observed under a large zenith 
angle, also indicate that the excess might continue along the 
spiral arm, possibly funneled by the magnetic fields which 
are believed to follow the spiral arm structure.

In the diffusion model we would also expect deficits in the 
direction out of the Galactic plane, since the Larmor
radius begins to exceed the thickness of the Galactic plane
(300\,pc) and particles can leak out.  This leakage is currently not 
observed, which could mean that there are strong magnetic 
turbulences which extend well out of the Galactic plane into 
the halo and prevent the leakage.
These halo fields have been observed in edge-on spiral galaxies
observed by radio astronomers, but for only a small
fraction ($5\,\%$) of edge-on galaxies~\cite{bib:hummel}.

\subsection{Neutral Particles}\label{ss:neutral}

Any model which assumes that charged particles are responsible 
for the observed enhancement predicts that the excess
around $10^{18}$\,eV should show substantial smearing, several
tens of degrees in diameter. 
While the excess observed by AGASA does indeed show substantial 
smearing, the SUGAR array observes an excess which is compatible
with a point source within the angular resolution of the array.
This may be an indication that the excess flux stems at least in
part from a neutral component, neutrons or $\gamma$-rays, which 
is not subject to deflection.  However, the absence of any 
anisotropy below $10^{18}$\,eV is hard to explain with 
$\gamma$-rays.  In addition, SUGAR, with its buried counters, 
is sensitive only to the muonic component of air showers, which 
rules out $\gamma$-ray primaries if our current understanding of 
air shower development is correct.  This leaves neutrons as a 
viable candidate.

Neutrons are produced by interactions between heavier nuclei and 
ambient photon fields in the source region, and by isospin flip 
in p-p-collisions.  In a remarkable ``coincidence,'' neutrons of 
energies around $10^{18}$\,eV have a $\gamma$-factor of $10^{9}$ 
and therefore a decay length which is close to the distance 
between us and the Galactic center.  Since neutrons can propagate 
this distance without decay or deflection, they provide a natural 
explanation for the narrow energy range of the AGASA 
observations: below about $10^{18}$\,eV, neutrons decay before 
reaching the Earth, and above $10^{18.4}$\,eV, the acceleration 
mechanism runs out of steam.

The observed energies of $10^{18}$\,eV cause a problem for both 
protons and neutrons.  In particular, as Biermann et 
al.~\cite{bib:biermann5} point out, the black hole at the 
center of our Galaxy, a natural candidate for a cosmic ray 
source, does not have sufficient activity to produce neutrons 
near $10^{18}$\,eV, and the same holds for mini-quasars in the 
Galactic center region.

\section{Scientific Goals}\label{s:goals}

The goal of \project is a detailed study of the cosmic ray 
flux in the region from $10^{17}$ to $10^{19}$\,eV. 
This includes
\begin{enumerate}
\item{a measurement of the energy spectrum and detailed study of spectral 
features like the ankle and the `second knee,'} 
\item{a measurement of the average chemical composition,} 
\item{a mapping of the Galactic center region, and a determination of 
the chemical composition and the energy spectrum in regions where an excess
is found.}
\end{enumerate}
The study of the flux and the composition will bridge the
energy gap between experiments like KASCADE~\cite{bib:kascade},
which studied the composition around the ``knee'' at 
$4\cdot 10^{15}$\,eV, and HiRes and Auger, which cover the range
from $10^{19}$\,eV up to the GZK cutoff or beyond.

The mapping of the Galactic center region needs to establish
where any excess originates, how many sources are responsible, 
and whether the source image shows smearing or appears to be 
point-like.
If indeed particular acceleration sites can be identified,
we need to measure their energy spectrum with high accuracy to 
learn about the acceleration mechanisms at work.  In particular,
the high energy end of the spectrum will provide valuable 
information on the conditions at the source (or sources).

The detector must be able to distinguish between $\gamma$-ray 
primaries, protons/neutrons, and heavier nuclei.  While most cosmic ray 
models are based on the acceleration of charged particles, other 
particle types inevitably occur after these primaries interact 
with the interstellar medium and produce (again) charged particles, 
neutrons, and neutral pions which decay into photons.  If SUGAR's 
excess is caused by neutrons, then the flux of cosmic rays in 
the source region and thus nucleon interaction must be extremely 
high, and a strong flux of $\gamma$-rays from pion decay must 
inevitably accompany the hadron flux.

These protons, neutrons, and photons provide important orthogonal 
information on Galactic sources of cosmic rays, as all three 
particle types probe different distances within the Galaxy.  
From {\it charged hadrons}, we expect a direct component only 
from a very limited region of space.  Only with neutrons near
$10^{18}$\,eV or with $\gamma$-rays can we study the Galactic center
region directly.  While $\gamma$-rays measure the line of sight 
integral through the entire disk and halo, neutrons decay after a 
decay length $d$, so we measure the integral up to a distance $d$.

The separation of heavier nuclei from protons and protons from 
$\gamma$-primaries is possible if the height of the shower maximum 
is known.  While the intrinsically large fluctuations of the shower
start and development make a separation on an event-by-event basis 
impossible, the average height of the shower maximum for an ensemble 
of showers is a strong indicator of the particle 
type.  Neutron- and proton-initiated air 
showers have identical signatures and can not be separated, but here,
the shape of the excess will provide important information.  Since 
charged particles can not reach us directly, any gradient in the
charged particle flux from the Galactic center region will cause a 
symmetric anisotropy with an excess from the direction of the center 
and a deficit from the anti-center.  


The most accurate technique today to determine the energy, the 
interaction characteristics, and the arrival direction of cosmic ray 
primary particles at ultra high energies around $10^{18}$\,eV is the
stereo air fluorescence technique.  This technique uses the Earth's
atmosphere as a large aperture calorimeter, where primary cosmic
ray particles interact with air molecules to produce huge cascades
of particles, so-called extensive air showers.  The particles of the
shower cascade excite and ionize air molecules which fluoresce in the 
UV.  The fluorescence light can be detected by photomultiplier cameras
watching the dark night sky, and a three-dimensional picture of the 
shower development in the atmosphere can be reconstructed from the 
measured trigger times and light intensitites observed by each tube
along the shower trajectory.  If the shower is observed by two detectors
simultaneously, the three-dimensional geometrical reconstruction 
of the shower axis is more precise, and systematic uncertainties due
to incomplete knowledge of atmospheric parameters are largely removed.

The air fluorescence technique was pioneered by the Fly's Eye 
collaboration, which operated a monocular detector from 1981 to 1986 
and a stereo detector from 1986 to 1993.  The HiRes experiment
is currently operating as a second-generation air fluorescence 
detector at the same site, on the Dugway Proving Grounds in Utah.

The air fluorescence technique requires clear dark nights and good
and stable atmospheric conditions.  The technique is therefore
limited to dry desert areas in remote locations with minimum light 
pollution.  These requirements severely reduce the duty cycle, to 
less than $10\,\%$.  In addition, constant monitoring of the 
atmosphere is necessary to correctly determine the aperture of the 
detector at any given time.  

These shortcomings are outweighed by decisive advantages over 
ground arrays.  A stereo air fluorescence detector can achieve 
angular resolution better than $1^{\circ}$, energy resolution of 
$20\,\%$, and can directly determine the position of the 
shower maximum to $20\,\mathrm{g\,cm}^{-2}$, an important requirement 
to determine the interaction characteristics and thus the type of 
the primary particle.

The design of \project is therefore based on the stereo air 
fluorescence technique as the most appropriate technique to achieve 
the scientific goals outlined in this section.

\begin{figure}
\begin{center}
\epsfig{file=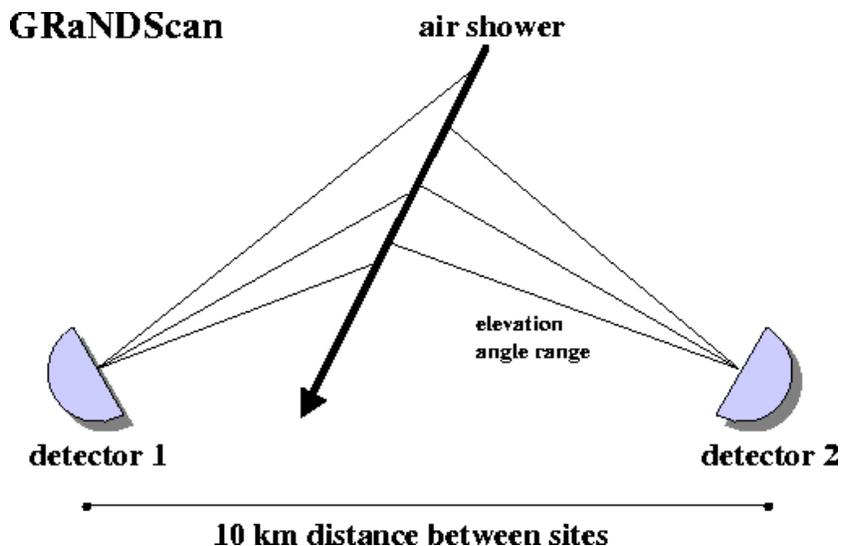,width=11cm}
\caption{\protect\sl Scheme of the \project detector geometry.}
\label{fig:detector}
\end{center}
\end{figure}

\section{\project Design}\label{s:design}

The design for \project has been driven by the attempt to
achieve a maximum aperture for stereoscopic observation
of air showers while keeping the number of telescopes small.
In the baseline design, \project is an air fluorescence detector 
with two sites about 12\,km apart which are facing each 
other and therefore view a common volume of atmosphere.  

A schematic of the detector is shown in Fig.\,\ref{fig:detector}.
Each site consists of several reflecting telescopes with a 
$3.4\,\mathrm{m}$ diameter mirror for light collection and
a photomultiplier camera in its focal plane. Each telescope
covers a field of view of $30^{\circ}$ by $30^{\circ}$.
The large field of view has been chosen to keep the number of
detector units small.  With 3 mirrors at each site, 
a range of $90^{\circ}$ in azimuth and $30^{\circ}$ in zenith 
angle can be covered.  The axis of the mirror has an angle to 
the horizontal of $50^{\circ}$.  The number of camera units, 
the pointing direction in zenith angle and the distance between 
the sites are chosen to maximize the aperture of the detector 
in the relevant energy region around $10^{18}$\,eV (see 
Section~\ref{ss:layout}).  Covering a larger range in azimuth 
and zenith increases the aperture, but at $10^{18}$\,eV, detected 
showers are mainly ``local,'' {\em i.e.} close to both detector 
sites.  Most of the sensitivity can be achieved with fewer units 
if we only consider the atmospheric volume in between the sites.

The configuration allows stereo observation of air showers 
and provides the angular resolution associated with this 
technique.  The need for a good angular resolution is evident 
in the case of extended sources, where the exact structure of 
the source needs to be studied.  However, a good angular 
resolution is also crucial for sensitivity to point 
sources.  

The figure of merit for a telescope which is aimed 
at the detection of a point-like source over a dominating 
isotropic background is the signal to noise ratio 
\begin{equation}
\left(\frac{\mathrm{signal}}{\mathrm{noise}}\right)
\propto\frac{R~Q~\sqrt{A_{eff}~T}}{\sigma_{\theta}}~~,
\label{eqn:signal}
\end{equation}
where $T$ is the exposure time, $A_{eff}$ is the effective 
detector area, and $\sigma_{\theta}$ is the angular resolution.  
The signal/noise relative trigger efficiency $R$ and the 
signal/noise identification efficiency $Q$ are close to 1 in 
our case since background and noise presumably have similar 
chemical composition and a separation of the signal from the 
noise is not possible. (However, $Q$ can be considerably larger 
than 1 for an excess produced by $\gamma$-ray primaries over a 
hadronic background).

In the following sections, we address how to optimize the
{\it exposure time} and the effective detector area or 
{\it aperture} of the detector.

\subsection{Location}\label{ss:location}

The experimental challenge for \project is to overcome an  
infrastructure problem.  Air fluorescence detectors require
sites without light pollution and with excellent atmospheric 
conditions,  which implies remote (desert) areas.
The costs for installing power lines to remote locations
is prohibitive.  However, with recent developments in low 
power electronics and large analog memories it is for the 
first time feasible to design and develop an air fluorescence 
camera that operates on solar power alone.  This means that 
only a minimal amount of infrastructure is required and
operating costs are kept at a minimum.  The minimal 
infrastructure also allows the detectors to be easily
moved or reconfigured.

\begin{figure}
\begin{center}
\epsfig{file=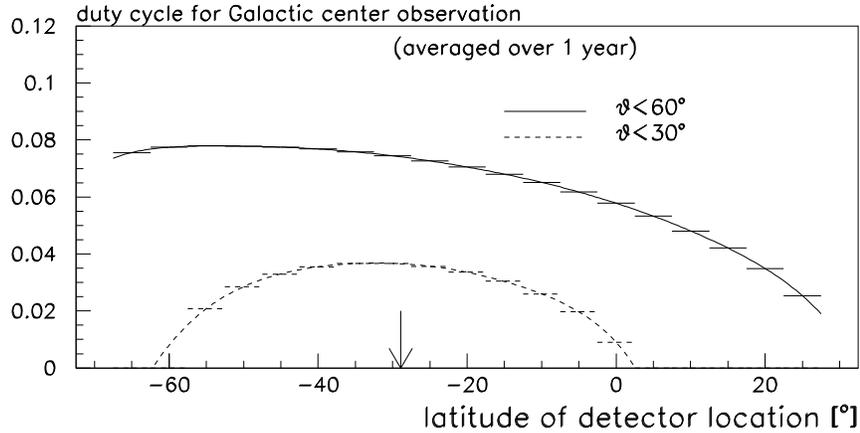,width=13cm}
\caption{\protect\sl The dutycycle for observation of the 
Galactic center as a function of detector latitude.  The solid 
line includes all times when the zenith angle $\theta$ of the 
Galactic center in local coordinates is less than $60^{\circ}$, 
the dotted line is for times with $\theta<30^{\circ}$.  The 
arrow indicates the declination of the Galactic center.}
\label{fig:dutycycle}
\end{center}
\end{figure}

Air fluorescence detectors have limited duty cycles since 
observation time is restricted to dark, moonless nights.
The duty cycle of an instrument which aims at the detection
of a specific source is even smaller, as the source can not 
be observed during the period of the year when it crosses 
the sky during daytime.

Neglecting downtime caused by bad weather, the observation 
time is determined by the geographical latitude of 
the detector location.  The Galactic center is at declination 
$\delta=-28.9^{\circ}$.  
A good visibility of this region therefore requires a detector 
location in the southern hemisphere.  Fig.\,\ref{fig:dutycycle}
shows the duty cycle as a function of the detector latitude 
integrated over a full year of observation.  The duty cycle 
is defined as the observation time divided by the total time 
(1 year), where observation time is defined as those time 
periods between moon set (or rise) and astronomical twilight 
where the Galactic center has a zenith angle of less than 
$60^{\circ}$ (solid line) and $30^{\circ}$ (dotted line) in 
local coordinates.  For a good visibility with small zenith
angles, a southern latitude corresponding to the declination
of the Galactic center optimizes the observation time.  If 
we allow the Galactic center to go down to $60^{\circ}$, 
more southern locations are equally well-suited, but it should
be stressed that at latitudes further south than $70^{\circ}$, 
the Sun starts to stay continuously above twilight limit for 
long periods of the year, and further south, it will also stay 
continuously below twilight limit. In our case, this is not 
an advantage, as we need sunlight periods to recharge batteries.  
An ideal site is therefore at moderate southern latitudes 
approximately equal to the declination of the Galactic
center where days are sufficiently long to allow daily recharging.
Duty cycles of $8\,\%$ ($4\,\%$) can be expected for
$\theta<60^{\circ}$ ($\theta<30^{\circ}$), but this number
is an upper limit, as it does not account for weather conditions.
For comparison, a ground detector array at a latitude of
$-30^{\circ}$ has a duty cycle for Galactic center observations
of $39\,\%$ ($19\,\%$).

For measurements of the energy spectrum and the average chemical
composition, which do not require the Galactic center to be in field
of view, the duty cycle is more favorable and reaches $xx\,\%$.

\begin{figure}
\begin{center}
\epsfig{file=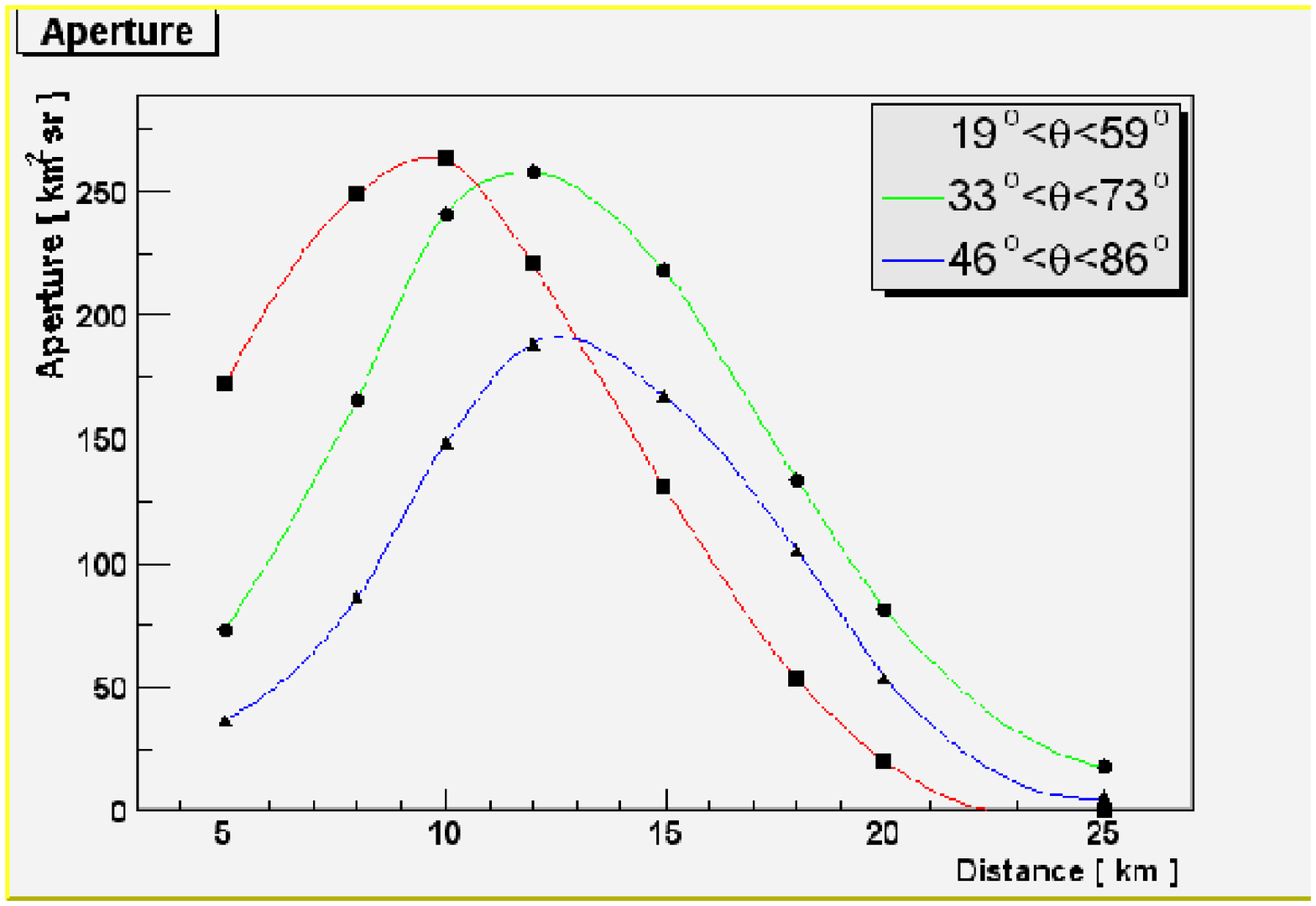,width=7cm}
\epsfig{file=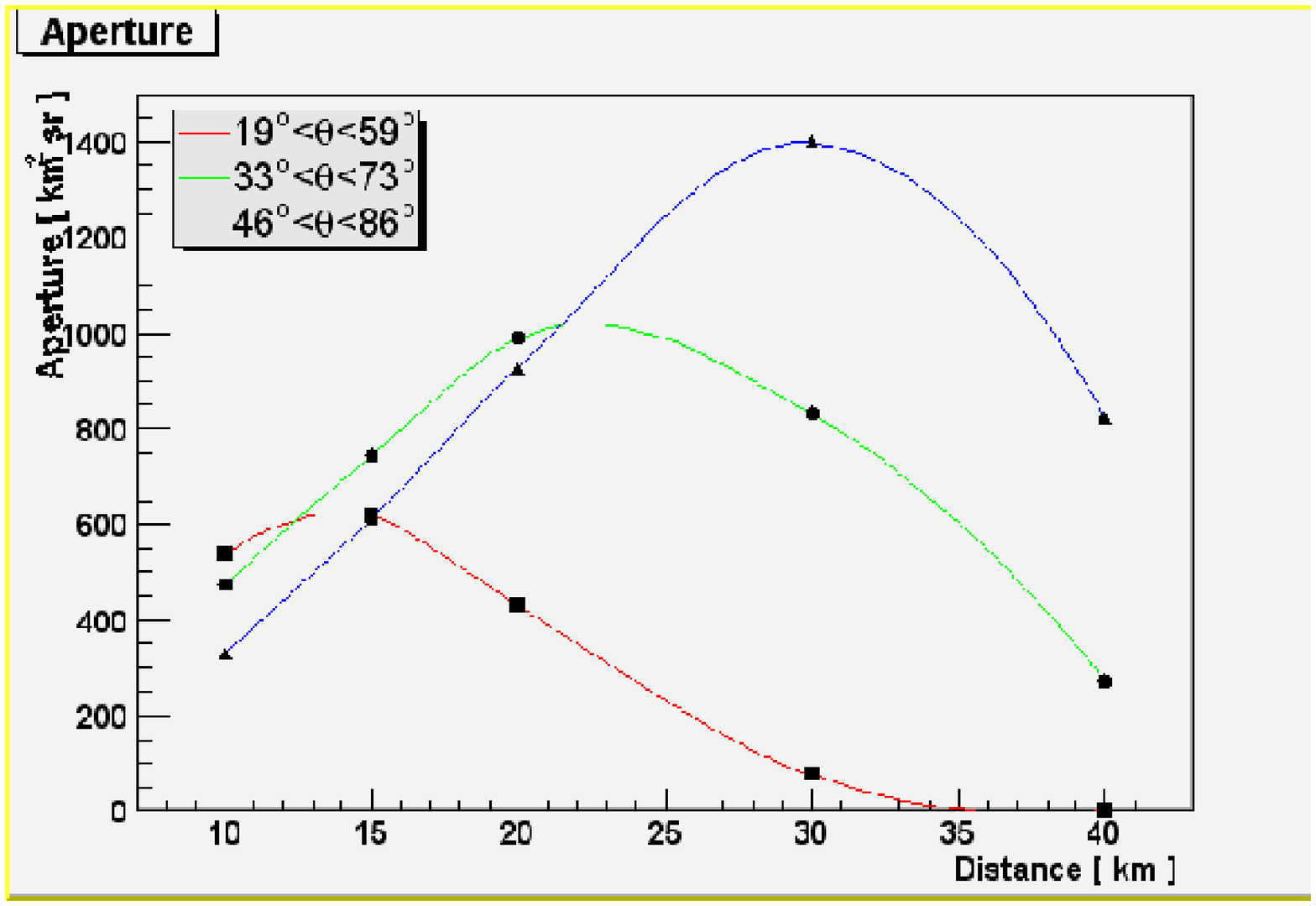,width=7cm}
\caption{\protect\sl Aperture as a function of distance between the
two detector sides for two different energies (left: $10^{18}$\,eV,
right: $10^{19}$\,eV).  The different colors indicate different zenith
angle ranges.}
\label{fig:aperture18}
\end{center}
\end{figure}

\subsection{Layout of the Sites}\label{ss:layout}

The distance of the two detector sites and the field of view
in local coordinates will determine the aperture of the
detector.  The optimal distance varies with shower energies.  
The atmospheric volume observed should be as large as possible, 
but if the distance between the sites is too large, only a 
small fraction of events are seen by both detectors.

Another parameter which determines the aperture is the zenith
angle range covered by the detector.  To keep the number of
mirror units small, the camera should not cover the whole
zenith angle range from $0^{\circ}$ to $90^{\circ}$, but only 
that fraction of the full range which contributes most to the
total aperture and is most likely to contain the shower maximum.
A reliable energy and shower maximum reconstruction requires 
that at least one of the sites observes the shower maximum in 
its field of view.

This optimal zenith angle range depends on the energy of interest.
High energy showers reach their maximum later, so the average
height of the shower maximum decreases with energy.  A detector
working at energies below $10^{18}$\,eV should therefore 
preferentially observe higher elevations, whereas at $10^{19}$\,eV 
and above, a lower elevation range will maximize the aperture.

To find the site layout which maximizes the aperture of the
stereo detector, we simulated the stereo aperture as a 
function of the distance between the sites and the zenith angle
viewing range.  To guarantee that the event can be reconstructed 
with good accuracy, an event is only accepted if each detector 
site has at least 5 signal tubes and the (reconstructed) shower 
maximum is in the field of view of at least one of the two sites.

Fig.\,\ref{fig:aperture18} shows the aperture as a function
of the detector distance for two different energies ($10^{18}$\,eV 
and $10^{19}$\,eV) and three zenith angle ranges.  As expected, both 
the optimal distance between the two sites and the optimal zenith 
angle range strongly depend on energy.  At higher energy, a larger 
distance and a field of view covering elevations closer to the 
horizon maximize the aperture.

The two energies illustrate a further advantage of a detector 
composed of movable units.  Optimal detector parameters change 
rather dramatically with each decade in energy.  It is 
conceivable that a mobile detector like \project can be operated 
at different site distances over its life time.  After an initial 
run at $10^{18}$\,eV, the detector distance can be changed to 
increase the aperture at higher or lower energies. 

\subsection{Photomultiplier Camera and Electronics}\label{ss:pmt}

A crucial element of \project is the light detector used in the 
air fluorescence camera.  At the lower end of the \project energy 
range, the signal to noise ratio will limit the lowest signal 
that can be measured.  This almost automatically leads to 
photomultiplier tubes as the best-suited camera element.  
Photomultipliers have typically high gain, single photon 
efficiency, low intrinsic noise, and do not require cooling,
and important advantage if power consumption is to be kept at
a minimum.

\begin{figure}
\begin{center}
\epsfig{file=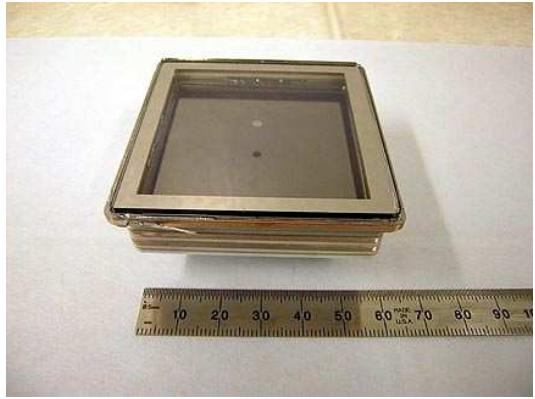,width=7cm}
\caption{\protect\sl The BURLE 85001 
$\mathrm{Planacon}^{\mathrm{TM}}$ multichannel plate 
photomultiplier.}
\label{burle}
\end{center}
\end{figure}

The \project camera needs to be light in weight for instrumenting
a remote detector.  A wide field  of view also requires a 
curved camera surface (see next section), which is difficult to
achieve with conventional photomultipliers.  Flat-panel 
multi-anode photomultipliers now overcome this limitation, and 
the anode pixellation of devices like the $2''$ BURLE 85001 
$\mathrm{Planacon}^{\mathrm{TM}}$~\cite{bib:burle} can be chosen
as to best match the application (see Fig.\,\ref{burle}).
2 by 2 anode configurations in a $2''$ phototube are readily 
available, which means each of these compact devices replaces 
four conventional photomultipliers.  A camera using flat
multi-anode photomultipliers will be considerably smaller and
less heavy than a traditional air fluorescence camera.

Current off-the-shelf solar power units deliver about 50\,W 
for 5 to 6 hours per $1\,\mathrm{m}^{2}$ paddle size.  
Therefore the limit on the power consumption per channel for 
an air fluorescence camera with a $30^{\circ}$ by 
$30^{\circ}$ field of view and just under 1000 photomultiplier 
channels running for 6 hours during a typical data taking 
night is 50 to 100\,mW per channel for one or two 
paddles.  This is roughly two orders of magnitude lower than 
the power consumption of a current HiRes camera.  However, 
these are numbers referring to the {\it average} power 
consumption.  Since a camera will ultimately consist of low 
power and high power elements, the goal is to keep the power 
consuming elements dormant most of the time and only activate 
them when an intelligent second-level trigger identifies a
shower candidate.  This trigger can be modeled after 
HiRes\,2, the FADC based second site of the HiRes 
detector~\cite{bib:hires_fadc} in Utah, where a possible 
scheme has been successfully implemented.  A Digital Signal 
Processor calculates the geometrical moments of the image 
formed by signal tubes on the photomultiplier camera and 
bases a trigger decision on the shape of this image.  The 
event is accepted if certain criteria for a track-like image 
are met.

\begin{figure}
\begin{center}
\epsfig{file=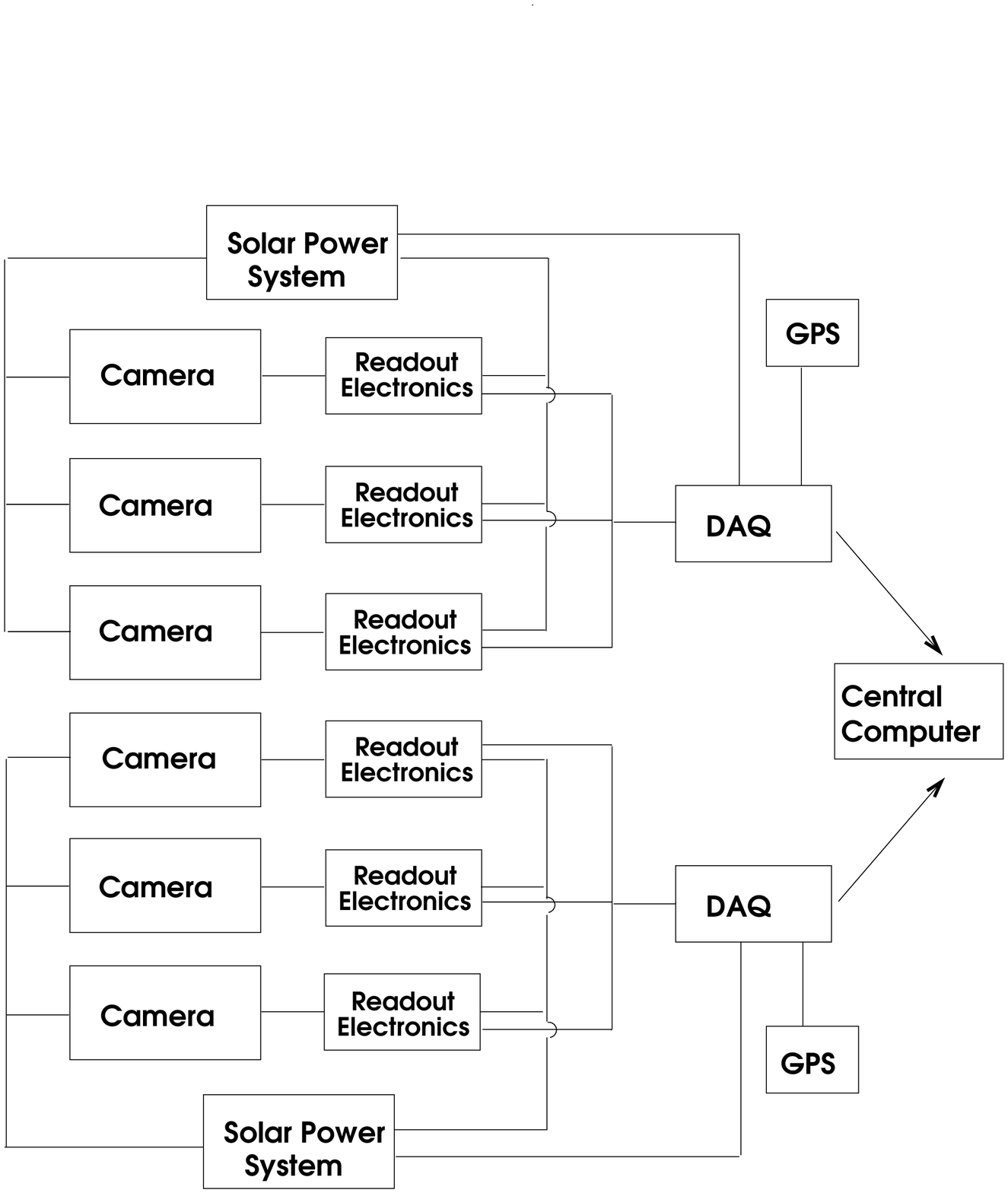,width=7cm}
\epsfig{file=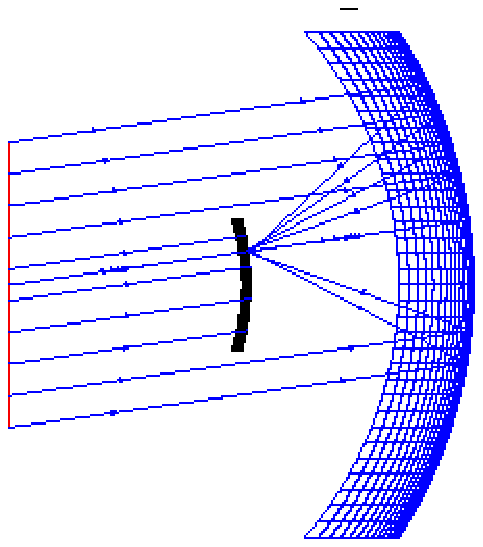,width=7cm}
\caption{\protect\sl (Left) Scheme of the detector. The two
sites are operated independently and are powered by a solar
panel. (Right) Wireframe scheme of a signal camera.  The radius
of curvature of the light collecting mirror is 3.4\,m.}
\label{camera_wireframe}
\end{center}
\end{figure}

While the trigger decision is pending, the data goes through 
an analog delay line (switched capacitor array).  For the expected 
signal from air showers at $10^{18}$\,eV, a sampling rate of 
10\,MHz and a dynamic range of 12 bits are appropriate.  
Both sites function as independent units with their own solar
power unit and central data acquisition combining the data
from the camera units.  A trigger is generated independently
from the other site and the data is broadcast to a central
station where the data sets are combined and matched.
Fig.\,\ref{camera_wireframe} (left) illustrates this scheme.
The data can be stored on hardware RAID systems, and the main
control of the experiment can be remote over the Internet,
linked by a satellite.

\subsection{Optical System}\label{s:optics}

The need to image extremely weak light tracks over an 
extended field of view of order $30^{\circ}$ by 
$30^{\circ}$ suggests a Schmidt optical system.  A Schmidt 
camera consists of a spherical mirror with a large aperture 
stop (e.g. $f/\#=1$), located at its center of curvature.
The mirror gives uniform images over a spherical surface 
concentric with itself, but the images suffer from 
spherical aberration, which in a classical optical Schmidt 
system are corrected with a refractive plate at the aperture.

\begin{table}
\begin{center}
\begin{tabular}{|l|r|}\hline
\multicolumn{2}{|c|}{\rule[-3mm]{0mm}{8mm}\bfseries TABLE 2 ---
      $\mathbf{15\times15}$ MCP Camera Parameters}\\ \hline
    Aperture diameter           &   1900.0~mm\\
    Mirror radius of curvature  &   3125.0~mm\\
    Mirror size                 &   3400.0~mm\\
    Camera radius of curvature  &   1622.2~mm\\
    Camera arc length           &    883.9~mm\\
    Anode pixelation            &   ($2\times2$) 25.4~mm\\
    Spot diameter               &     25.4~mm\\
    $1^{\circ}$ image size      &     28.3~mm\\
    Resolution                  &   $0.9^{\circ}$ per pixel\\
    Field of view               &   $30^{\circ}$\\
    Camera obscuration          &    26.9\% center, 24.4\% corner\\
\hline
\end{tabular}
\caption{Multichannel plate camera parameters.}
\label{table1}
\end{center}
\end{table}

The resolution of a photomultiplier camera is determined by 
the pixel size of the photomultiplier. Consequently, a 
reduction of the image spot to a dimension much smaller than 
the pixel is not necessary.  Simulations using the 
ZEMAX~\cite{bib:zemax} ray tracing code show that
a Schmidt system without a corrector plate is 
therefore sufficient if the field of view is not much larger 
than $30^{\circ}$ by $30^{\circ}$.  For a given mirror radius 
of curvature radius $R$, the diameter $D$ of the aperture stop 
largely determines the size of the circular image spot.  In 
considering various geometries, we have been guided by the 
following requirements:
\begin{enumerate}
\item{a $30^{\circ}$ by $30^{\circ}$ field of view,}
\item{$1^{\circ}$ per pixel resolution or better,}
\item{the size of the image spot be roughly the pixel size,}
\item{the camera obscures the mirror by not more than
$25\%$.}
\end{enumerate}
Calculations using ZEMAX show that an optical system meeting 
these requirements is feasible.  Table\,\ref{table1} summarizes 
the dimensions of a $15\times15$ microchannel plate PMT camera, 
Fig.\,\ref{camera_wireframe} shows a wire frame schematic of the 
camera design.  The aperture diameter is $1900$\,mm and the 
mirror size is $3400$\,mm.  The image spot diameter of 
$25.4$\,mm matches the pixel size of the BURLE 85001 tube 
with a $2\times2$ anode pixellation.  
It should be noted that these optimization results are similar
to the design of the air fluorescence cameras of the 
Auger detector~\cite{bib:mh99,bib:gm01}, although the final
Auger design includes a partial corrector plate at the
aperture stop.

For a larger field of view, for example $40^{\circ}$ by 
$40^{\circ}$, a corrector plate becomes necessary. 
The Orbiting Wide-Angle Light Collectors (OWL) experiment
for example proposes a $45^{\circ}$ Schmidt camera with corrector
plate~\cite{bib:owl}.  Ultimately, the cost of a corrector plate 
has to be weighted against the cost of additional camera units 
if the desired field of view has to be covered by a larger number 
of smaller telescopes.

\subsection{Sensitivity to a Galactic Center Source}\label{ss:sensitivity}

The only flux measurement so far is based on the excess seen in
the SUGAR data.   
To estimate the sensitivity of \project to a flux of this magnitude, 
we calculate the aperture required to observe the excess at a 
significance $>10\,\sigma$ is {\em one year of data taking}.  
For this calculation, the SUGAR excess is treated as a signal 
coming from a point source.
The angular resolution and the aperture of \project have to compensate
for the reduced duty cycle of an air fluorescence detector.
If we estimate (conservatively) that the angular resolution
will improve by a factor of 3 over SUGAR, then \project needs
an increase in aperture of a factor of 150 over SUGAR.  
Fig.\,\ref{fig:aperture18} shows that this can easily be achieved with a
detector of moderate size.  

\subsection*{Acknowledgements}

We want to thank Peter Biermann, Roger Clay, Bruce Dawson, and Cyrus Hoffman
for fruitful discussions and valuable comments on the manuscript. We also
benefited from discussions with Frank Jones and Minghuey A.\,Huang.\\
The development of a solar-powered air fluorescence camera is 
supported by the National Science Foundation under grant number
NSF-PHY-0134007.


\pagebreak

\end{document}